\documentstyle[12pt,epsf]{article}
\expandafter\ifx\csname amssym12.def\endcsname\relax \else\endinput\fi
\expandafter\edef\csname amssym12.def\endcsname{%
       \catcode`\noexpand\@=\the\catcode`\@\space}
\catcode`\@=11
\def\undefine#1{\let#1\undefined}
\def\newsymbol#1#2#3#4#5{\let\next@\relax
 \ifnum#2=\@ne\let\next@\msafam@\else
 \ifnum#2=\tw@\let\next@\msbfam@\fi\fi
 \mathchardef#1="#3\next@#4#5}
\def\mathhexbox@#1#2#3{\relax
 \ifmmode\mathpalette{}{\m@th\mathchar"#1#2#3}%
 \else\leavevmode\hbox{$\m@th\mathchar"#1#2#3$}\fi}
\def\hexnumber@#1{\ifcase#1 0\or 1\or 2\or 3\or 4\or 5\or 6\or 7\or 8\or
 9\or A\or B\or C\or D\or E\or F\fi}
\font\tenmsa=msam10 scaled\magstep1
\font\sevenmsa=msam7 scaled\magstep1
\font\fivemsa=msam5 scaled\magstep1
\newfam\msafam
\textfont\msafam=\tenmsa
\scriptfont\msafam=\sevenmsa
\scriptscriptfont\msafam=\fivemsa
\edef\msafam@{\hexnumber@\msafam}
\mathchardef\dabar@"0\msafam@39
\def\dashrightarrow{\mathrel{\dabar@\dabar@\mathchar"0\msafam@4B}}
\def\dashleftarrow{\mathrel{\mathchar"0\msafam@4C\dabar@\dabar@}}

\def\ulcorner{\delimiter"4\msafam@70\msafam@70 }
\def\urcorner{\delimiter"5\msafam@71\msafam@71 }
\def\llcorner{\delimiter"4\msafam@78\msafam@78 }
\def\lrcorner{\delimiter"5\msafam@79\msafam@79 }
\def\yen{{\mathhexbox@\msafam@55 }}
\def\checkmark{{\mathhexbox@\msafam@58 }}
\def\circledR{{\mathhexbox@\msafam@72 }}
\def\maltese{{\mathhexbox@\msafam@7A }}
\font\tenmsb=msbm10 scaled\magstep1
\font\sevenmsb=msbm7 scaled\magstep1
\font\fivemsb=msbm5 scaled\magstep1
\newfam\msbfam
\textfont\msbfam=\tenmsb
\scriptfont\msbfam=\sevenmsb
\scriptscriptfont\msbfam=\fivemsb
\edef\msbfam@{\hexnumber@\msbfam}

\def\widehat#1{\setbox\z@\hbox{$\m@th#1$}%
 \ifdim\wd\z@>\tw@ em\mathaccent"0\msbfam@5B{#1}%
 \else\mathaccent"0362{#1}\fi}
\def\widetilde#1{\setbox\z@\hbox{$\m@th#1$}%
 \ifdim\wd\z@>\tw@ em\mathaccent"0\msbfam@5D{#1}%
 \else\mathaccent"0365{#1}\fi}
\font\teneufm=eufm10 scaled\magstep1
\font\seveneufm=eufm7 scaled\magstep1
\font\fiveeufm=eufm5 scaled\magstep1
\newfam\eufmfam
\textfont\eufmfam=\teneufm
\scriptfont\eufmfam=\seveneufm
\scriptscriptfont\eufmfam=\fiveeufm

\csname amssym.def\endcsname
\newif{\ifcomentarios}
\comentariosfalse
\renewcommand{\theequation}{\thesection.\arabic{equation}}

\newcommand{\zerarcounters}
{
\setcounter{equation}{0}
\setcounter{theorem}{0}
}

\newcommand{\be}{\begin{equation}}
\newcommand{\ee}{\end{equation}}
\newcommand{\bma}{\begin{displaymath}}
\newcommand{\ema}{\end{displaymath}}
\newcommand{\bc}{\begin{center}}
\newcommand{\ec}{\end{center}}
\newcommand{\text}{\rm}

\newcommand{\uflex}
{{\scriptstyle {\raise 9pt\hbox{$\backslash$}\,\!\!\!\!\!\Bigg\vert}}}
\newcommand{\ncm}{\newcommand}

\ncm{\rncm}{\renewcommand}
\ncm{\id}{{\bf 1}}
\ncm{\beq}{\begin{equation}}
\ncm{\eeq}{\end{equation}}
\ncm{\ba}{\begin{array}}
\ncm{\bea}{\begin{eqnarray}}
\ncm{\beanon}{\begin{eqnarray*}}
\ncm{\ea}{\end{array}}
\ncm{\eea}{\end{eqnarray}}
\ncm{\eeanon}{\end{eqnarray*}}
\ncm{\fns}{\footnotesize}
\ncm{\setc}[1]{\setcounter{equation}{#1}}
\newcounter{eqnr}
\newenvironment{eqnarrayabc}{\stepcounter{equation}
  \setcounter{eqnr}{\value{equation}}\setc{0}
  \rncm{\theequation}{\thesection.\arabic{eqnr}\alph{equation}}
  \begin{eqnarray}}{\end{eqnarray}\setc{\value{eqnr}}}
\ncm{\eqboxabc}[3]{\newline\parbox[t]{1.5cm}{#1}\hfill
  \parbox[b]{12cm}{\begin{eqnarray*} #3\end{eqnarray*}}\hfill
   \parbox[b]{1.5cm}{\vspace{-0.0cm}\begin{eqnarrayabc}#2\end{eqnarrayabc}}\newline}
\ncm{\eqbox}[2]{\newline\parbox{1.5cm}{#1}\hfill
  \parbox{12cm}{\beanon #2\eeanon}\hfill
  \parbox{1cm}{\bea\eea}\newline}
\ncm{\nr}[1]{\parbox{1cm}{\begin{eqnarrayabc}#1\end{eqnarrayabc}}\\}

\ncm{\kal}[1]{\mbox{$\cal #1 $}}
\ncm{\mrk}[1]{\!\!\! #1 \!\!\!} 
\ncm{\qed}{\hspace*{0.4cm}\rule{0.24cm}{0.24cm}}  
\ncm{\mbold}[1]{\mbox{\boldmath $ #1 $}}   
\ncm{\bm}{\mbold}
\ncm{\str}{\stackrel}
\ncm{\sub}{\subset}
\ncm{\e}{\varepsilon}
\ncm{\ka}{\kappa}
\ncm{\inputc}[1]{\begin{center}\input{#1}\end{center}}
\ncm{\lto}{\longrightarrow}
\ncm{\x}{\times}
\ncm{\bmm}{\bm{\cal M}}
\ncm{\cp}{{\bf P}}    
\ncm{\bfp}{{\bf P}}
\ncm{\bmi}{\bm{i}}
\ncm{\bmom}{\bm{\om}}
\ncm{\bmOm}{\bm{\Om}}
\ncm{\res}{\restriction}
\ncm{\bmL}{\bm{\cal L}}
\ncm{\bmell}{\bm{\ell}}
\ncm{\bmE}{\bm{\cal E}}
\ncm{\bme}{\bm{e}}
\ncm{\bmpi}{\bm{\pi}}
\ncm{\bmr}{\bm{r}}
\ncm{\bmsigma}{\bm{\sigma}}
\ncm{\wt}{\widetilde}
\newcommand{\beaa}{\begin{eqnarray}}
\newcommand{\eeaa}{\end{eqnarray}}

\begin{document}

\author{{\bf Oscar Bolina}\thanks{Supported by FAPESP under grant
97/14430-2 and CNPQ. {\bf E-mail:} bolina@lobata.math.ucdavis.edu} \\
Department of Mathematics\\
University of California, Davis\\
Davis, CA 95616-8633, USA\\
\and {\bf J. Rodrigo Parreira}\\
Instituto de Estudos Avan\c cados\\
Rua Bar\~ao do Triunfo 375/304\\
04602-000 S\~ao Paulo, Brasil
}
\title{\vspace{-1in}
{\bf Bounds on Correlation Functions of Quantum Rotators}}
\date{\footnotesize{Received 16 September 1998}}
\maketitle
\begin{abstract}
\noindent
We derive a McBryan-Spencer bound to the correlation function of a
one-dimensional array of quantum rotators in the Villain
approximation of the cosine interaction. We obtain the partition
function of the system in the gas representation and establish a lower
bound on the external charges correlation function. We also discuss
the possible existence of a Kosterlitz-Thouless phase for the quantum
rotators in the Villain approximation.

\noindent
{\bf Key words:} Quantum Rotators, Villain Action, Correlation
Estimates, Disorder Operator \hfill \break
{\bf PACS Numbers:} 05.50.+q, 05.30.-d, 05.20.-y, 04.60.Nc
\end{abstract}



\section{Introduction}
\zerarcounters

We study the ground state of a system of coupled quantum rotators in
a one-dimensional lattice $\Lambda \subset {\bf Z}$ given by
the finite-volume Hamiltonian 
\beq
H_{\Lambda}=\frac{1}{2I} \sum_{x \in \Lambda} 
\frac{{\partial}^{2}}{\partial \theta^{2}} 
+ J \sum_{x \in \Lambda} \cos(\theta(x)-\theta(x+1)) \label{HM}
\eeq
$x \in {\bf Z}$, where {\it I} and $(J >0)$ are constant.
\newline
The Hamiltonian operator, taken with periodic boundary conditions,   
acts on the Hilbert space of square integrable functions on the interval
$[-\pi,\pi]$.
\newline
We use the Lie-Trotter representation to map our one-dimensional
model into a two-dimensional system of classical rotators with an extra
time direction. For this we prove a McBryan-Spencer bound on the
correlation function in the Villain approximation of the cosine
interaction (\ref{HM}).
\newline
Next we use a duality transformation \cite{FS} to obtain the partition
function in the {\it sine}-Gordon and charge representations,
which is a Villain gas partition function with anisotropic Gaussian
measure.
\newline   
We establish a lower bound on the external charges correlation
function through Jensen inequality in the charge variables.
\newline
Our estimates are for the infinite system in the continuum limit in
time direction.


\section{The Lie-Trotter Representation}
\zerarcounters

We derive a path integral representation for $Tr e^{-\beta H_{\Lambda}}$
by the Lie-Trotter formula
\beq
e^{-\beta H_{\Lambda}}=\lim_{n \rightarrow \infty} 
(e^{-\frac{\beta H_{0}}{n}}e^{-\frac{\beta V(\theta)}{n}})^{n} \label{TR}
\eeq
where
\[
H_{0}=\frac{1}{2I}\sum_{x}\frac{{\partial}^{2}}{\partial
\theta^{2}} \;\;\; {\rm and\ } \;\;\; V=J\sum_{x}
\cos[\theta(x)-\theta(x+1)]
\]
In computing the trace we take a basis that diagonalizes {\it V}
and insert between each two factors of (\ref{TR}) the decomposition 
of the identity $I=\sum_{\theta} |\theta(x)\rangle \langle \theta(x)|$
with $| \theta(x) \rangle$ eigenvectors of {\it V}.  
We get \cite{KP}
\beq
{\cal Z}=\lim_{n \rightarrow \infty} \int_{-\pi}^{\pi} 
e^{-H_{\Lambda,\delta}} \prod_{x,t} d\theta(x,t) \label{PF}
\eeq
where $e^{-H_{\Lambda,\delta}}$, with $\delta={\beta}/{n}$, is 
the Gibbs weight of a configuration $\theta=\{ \theta(x,t), x \in
\Lambda, t \in \delta{\bf Z} \cap [-{\beta}/{2}, {\beta}/{2}] \}$,
given by
\[
e^{-\beta H_{\Lambda},\delta}=\sum_{{\bf m_{1}}} e^{-\frac{I}{2 \delta} 
\sum_{x,t} (\theta(x,t)-\theta(x,t+\delta)+2 \pi m_{1}(x,t))^{2}} 
e^{-\frac{\delta J}{2} \sum_{x,t} \cos(\theta(x,t)-\theta(x+1,t))}
\]
where ${\bf m_{1}}=\{ m_{1}(x,t), x \in \Lambda, t \in \delta{\bf Z} 
\cap [-{\beta}/{2}, {\beta}/{2}] \}$. 
\newline
We want to replace the cosine form of the interaction by  
the Villain form \cite{V} 
\[
e^{z\cos{\theta}} \approx \sum_{m \in Z} 
e^{-\frac{z}{2}(\theta+2\pi m)^{2}}
\]
which is valid for $z \gg 1$.
\newline
In our case $z={\delta J}/{2}$ violates that condition. However,
we change our original model and define a new one for which the 
Gibbs weight {\it is} taken with the Villain action: 
\begin{equation}
e^{-H_{\Lambda,\delta}} = \sum_{{\bf m_{1}}, {\bf m_{2}}} 
\prod_{x,t} e^{- \{\frac{I}{2 \delta} [\theta(x,t)-\theta(x,t+\delta)
+2 \pi m(x,t)]^{2} + \frac{\delta J}{2} [\theta(x,t)-\theta(x+1,t) 
+ 2 \pi m_{2}(x,t)]^{2} \} }  \label{G}
\end{equation}
where ${\bf m_{2}}=\{ m_{2}(x,t), x \in \Lambda, t \in \delta{\bf Z} 
\cap [-{\beta}/{2}, {\beta}/{2}] \}$. \\
\newline
{\footnotesize {\bf Remark 1.} In forcibly approaching our model to a
classical lattice Villain model we have in mind proving a
Kosterlitz-Thouless transition for it. Since classical lattice and also
continuous Coulomb gases \cite{FS,DH} exhibit a Kosterlitz-Thouless
phase, it seems natural to expect our {\it anisotropic} Villain gas to
display such a phase as well. 
\newline
The McBryan-Spencer bound we derive below guarantees that our model does
not have a first order transition, but says nothing about transitions of
higher order.}  \\
\newline 
We are now in position to obtain a McBryan-Spencer bound on the 
correlation function. 
\newline
{\bf Theorem 1.} Let 
\beq
G_{\Lambda,\delta}((y,s),(y',s'))=
\langle  e^{\imath(\theta(y,s)-\theta(y',s')}
\rangle_{\Lambda,\delta} \label{MBS}
\eeq
where $\langle  \cdot \rangle_{\Lambda,\delta}$ is the mean value with
respect to the Gibbs weight (\ref{G}) for all $\Lambda, \delta$.
\newline
Let $G((y,s),(y',s'))$ be the limit of  
$G_{\Lambda,\delta}((y,s),(y',s'))$ when $\Lambda, n, \beta 
\rightarrow \infty$.
\newline
Then
\[
G((y,s),(y',s')) \leq
e^{-\sqrt{\frac{IJ}{4\pi^{2}}} \sqrt{(y-y')^{2}+(J/I)(s-s')^{2}}} 
\]
{\bf Proof.} Let $a(x,t)$ be a $C^{(2)}$-function. We make the shift
$\theta(x,t) \rightarrow \theta(x,t) + a(x,t)$, where 
\[
a(x,t)= C(x-y,t-s)-C(x-y',t-s'),
\]
and $C(x,t)$ is the Green's function of the finite difference
Laplacian in two dimensions, given by
\beq
C(x-y, t-s)=\frac{1}{| \Lambda |} \frac{1}{\beta} \sum_{p,q}
\frac{\exp{\imath (p(x-y)+q(t-s))}}
{J(2-2\cos{p}) + I{\delta}^{-2} (2-2\cos{q\delta}) }
\label{FG}
\eeq
with $p={2 \pi}/{| \Lambda |} \cap [-{|\Lambda|}/2,{|\Lambda|}/2] \;\;\;
{\rm and\ }  \;\;\; q={2 \pi}/{\beta} \cap [-n/2,n/2]$.
\newline
We obtain a quadratic form
\beq
\langle e^{\imath(\theta(y,s)-\theta(y',s')}\rangle \leq 
e^{-(a(y,s)-a(y',s'))}
e^{-\frac{\delta}{2} ( a, -\Delta a ) } \label{QDR}
\eeq
where $ -\Delta =J \partial^{*}_{1}\partial_{1} +
I \partial^{*}_{\delta}\partial_{\delta} $, with the usual definitions 
$\partial_{\delta}f(x,t)=\delta^{-1}[f(x,t+\delta)-f(x,t)]$, and scalar
product $(f,g)=\sum_{x,t}f(x,t)g(x,t)$.
\newline
Thus we get from (\ref{QDR})
\beq
\langle e^{\imath(\theta(y,s) - \theta(y',s'))} \rangle_{\Lambda,\delta} 
\leq e^{(C(y-y',s-s')-C(0,0))} \label{LL}
\eeq
The asymptotic behavior of the difference of Green's functions in
(\ref{LL}) in the limit $\Lambda, n , \beta \rightarrow \infty$ is:
\[
C(y-y',s-s') - C(0,0) \approx - \sqrt{\frac{1}{4\pi^{2}IJ}} 
\ln{\sqrt{(y-y')^{2}+(J/I) (s-s')^{2}}}
\]
for large $|y-y'|$, $|s-s'|$. 
\newline
It follows that 
\[
\;\;\;\;\;\;\;\;\;\; \langle e^{\imath(\theta(y,s) - \theta(y',s')}
\rangle \leq e^{-\sqrt{{IJ}/{4\pi^{2}}}~ \ln{\sqrt{(y-y')^{2}+
(J/I) (s-s')^{2}}}} \;\;\;\;\;\;\;\;\Box
\]
\noindent
This should be compared with a similar result in \cite{KP}.


\section{Duality Transformation}
\zerarcounters

Duality transformation is the representation of the partition function 
obtained by performing the Fourier transform on its angle variables 
$\{ \theta(x,t) \}$. For the partition function (\ref{PF}) with the
Gibbs weight (\ref{G}) the transformation is trivial. After duality
the partition function reads 
\beq
{\cal Z}=\int \prod_{x,t} \sum_{m \epsilon {\bf Z}}
\delta(\phi(x,t)-m(x,t)) 
d\mu(\phi)  \label{SG}
\eeq
where $\{ \phi(x,t) \}$ are the variables of the dual lattice, and
$d{\mu}(\phi)$ is the discrete Gaussian measure
\begin{equation}
\prod_{x,t} e^{-\frac{\delta}{2I} (\phi(x,t)-\phi(x+1,t))^{2}
-\frac{\delta}{2J} (\frac{\phi(x,t)-\phi(x,t+\delta)}{\delta})^{2}}
d\phi(x,t) \label{GAU}
\end{equation} 
The periodized $\delta$-functions can be expanded in Fourier 
series 
\[
\prod_{x,t} \{ \sum_{m(x,t)} \delta(\phi(x,t)-m(x,t)) \}=
\sum_{q} e^{\imath 2 \pi (\phi,q)} 
\]
where $q=\{ q(x,t) \in {\bf Z} \}$, and we get
\beq
{\cal Z}=\int d{\mu}(\phi) \prod_{x,t} (1+2\sum_{q=1}^{\infty}
\cos{2 \pi \phi(x,t)q(x,t)}). \label{C}
\eeq
On integrating out the $\phi$-variables in (\ref{C}) we obtain the
charge representation of the partition function:  
\beq
{\cal Z}= \int e^{-\frac{1}{2}(q, (-\Delta)^{-1} q)} \prod_{x,t}
d\lambda(q(x,t))  \label{Q}
\eeq
where now $ -\Delta =I^{-1}\partial^{*}_{1}\partial_{1} +
J^{-1} \partial^{*}_{\delta}\partial_{\delta} $, with the measure
\[
d\lambda(q)=\sum_{m \epsilon {\bf Z}} \delta(q - 2 \pi m) dq
\]
If we put into the system (\ref{Q}) two fractional charges, one $+\xi$
placed at the origin, another $-\xi$ placed at {\it x}, the external
charges correlation function reads
\beq
G^{\xi}_{\Lambda,\delta}(x)={\cal Z}^{-1}
\int e^{-\frac{1}{2}(q + \rho(y,t), -\Delta^{-1}(q +\rho(y,t)))} 
\prod_{y,t} d{\lambda(q(y,t))} \label{CF}
\eeq
where $\rho(y,t)=\xi(\delta_{y0}-\delta_{yx})\delta_{t0}$ is the 
density of external charges. 
\newline
Let $G^{\xi}(x)$ denote the limit of $G^{\xi}_{\Lambda,\delta}(x)$ 
when $\Lambda, n, \beta \rightarrow \infty$.  We have
\newline
{\bf Theorem 2.} The external charges correlation function (\ref{CF}) 
is bounded below by
\[
G^{\xi}(x) \geq C_{\xi} e^{-\xi^{2}\sqrt{\frac{IJ}{4\pi^{2}}}\ln{|x|}}
\]
with $0< C_{\xi} < \infty.$ 
\newline
{\bf Proof.} Immediate by the application of Jensen inequality in 
the $q$  -- variables.


\section{The Disorder Operator}
\zerarcounters

When the duality transformation is applied to the correlation function
(\ref{MBS}), its expectation value for the correlation between the points
$0$ and $(x,0)$ can be written as $G_{\Lambda,\delta}(0,x)=\langle
D^{\xi}_{0x} \rangle^{\phi}$ where $\langle D^{\xi}_{0x} \rangle^{\phi}$
is the expectation value, taken with the measure (\ref{GAU}), of the
so-called disorder operator 
\beq
\hat{D}^{\xi}_{0x}(\phi)=\frac{v(\partial_{2} \phi + \xi f^{x})}
{v(\partial_{2} \phi)} \label{DI} 
\eeq
where
\beq
v(\partial_{2} \phi)=e^{-\frac{\delta}{2J} \sum_{y,t}(
\frac{\phi(y,t)-\phi(y,t+\delta)}{\delta})^{2}} \label{VVV}
\eeq
and $f^{x}$ is defined by   
\begin{equation} f^{x}(y, t)=\cases{1,&if $1\leq y \leq x,
\;\; 
t=0$; \cr
                            0,&otherwise.\cr} \label{6.14}
\end{equation}
From (\ref{DI}) and (\ref{C}) we get
\[
{\cal Z}\langle\hat{D}^{\xi}_{0x} \rangle^{\phi} = e^{-\frac{\delta}{2J}
{\xi}^{2}|x|} \int d{\mu}(\phi) e^{\frac{\delta}{J} \xi (\phi,
\partial_{\delta} f^{x})}  \prod_{y,t} \left (1+2\sum_{q=1}^{\infty}
\cos{2 \pi \phi(y,t)q(y,t)} \right )
\]
The real shift $\phi \rightarrow \phi + \sigma$,
where $\sigma(y,t)=({\xi}/J)(-\Delta^{-1})\partial_{\delta}
f^{x}(y,t)$ leads to
\[
{\cal Z}\langle\hat{D}^{\xi}_{0x} \rangle^{\phi} = e^{g(x)} Z(\sigma)
\]
where
\beq
Z(\sigma)= \int \prod_{y,t} 
\left [1+2\sum_{q=1}^{\infty}\cos(2 \pi (\phi(y,t)+\sigma(y,t))q(y,t))
\right ] d{\mu}(\phi)  \label{FINAL} 
\eeq
and 
\[
g(x)=-\frac{\delta \xi^{2}}{IJ} (C(0,0)-C(x,0))
\]
is the Gaussian contribution to the expectation value, which vanishes
in the limit when $\Lambda, n, \beta \rightarrow \infty$.\\ 
\newline
{\footnotesize {\bf Remark 2.} The application of Jensen inequality to
obtain the decay of the correlation function is the first step towards
proving a Kosterlitz-Thouless transition for the quantum rotators. The
second step consists in obtaining a lower bound to the expectation value
of the disorder operator. Starting from (\ref{FINAL}) expectations in
the gas are written as convex combinations of expectations of dilute gases
of neutral multipoles. The falloff of correlation intended to imply the
transition is obtained by an inductive procedure over multipoles of
various sizes. In each step of the process the entropy of the multipoles
increases by a factor proportional to the diameter of the support of
charges. This entropy increase must be offset by the extraction of the
self-energy of charged multipoles in order to renormalize the activity 
of the gas. The energy estimate yields roughly a factor $e^{-C(0,0)}$.
\newline
Naive application of these techniques to the {\it partition function}
(\ref{FINAL}) shows that the entropic cost grows as the inverse of the
spacing $\delta^{-1}$, while the self-energy factor grows as
$\exp{\delta}$, so that a more careful analysis is required to render the
balance energy-entropy favorable. }


\end{document}